\documentclass[11pt]{article}

\usepackage{fullpage,latexsym,amsmath,amsfonts,color,epsfig}
\usepackage[latin1]{inputenc} 

\newcommand{\bfT}{{\bf T}}

\newcommand{\calX}{{\cal X}}
\newcommand{\calY}{{\cal Y}}

\newcommand{\pmtn}{{\mbox{\rm pmtn}}}

\newcommand{\onehalf}{{\mbox{$\frac{1}{2}$}}}

\newcommand{\braced}[1]{{ \left\{ #1 \right\} }}

\DeclareMathSymbol{\reals}{\mathbin}{AMSb}{"52}

\newcommand{\calXbef}[1]{\calX_{<#1}}
\newcommand{\calXaft}[1]{\calX_{>#1}}

\newcommand{\undefined}{{\bot}}

\newtheorem{theorem}{Theorem}

\newtheorem{lemma}[theorem]{Lemma}

\newtheorem{claim}[theorem]{Claim}

\newenvironment{proof}{{\it Proof:\/}}{$\Box$\vskip 0.1in}

                {$\spadesuit$\smallskip}

\newenvironment{bigeqn*}{\large\begin{eqnarray*}}{\end{eqnarray*}}

\newcommand{\SetFigFont}[6]{#6}

\begin{document}


\title{Preemptive Multi-Machine Scheduling of Equal-Length Jobs \\
        to Minimize the Average Flow Time} 

\author{
Philippe Baptiste%
\thanks{CNRS LIX, Ecole Polytechnique,
91128 Palaiseau, France.
baptiste@lix.polytechnique.fr.}
\and
Marek Chrobak%
\thanks{Department of Computer Science,
University of California,
Riverside, CA 92521.
marek@cs.ucr.edu.
Supported by NSF grants CCR-0208856 and INT-0340752.
}
\and
Christoph Dürr%
\thanks{LRI UMR 8623,
Université Paris-Sud,
91405 Orsay, France.
durr@lri.fr.
Supported by the EU 5th framework programs RESQ IST-2001-37559, the
NSF/CNRS grant 17171 and the CNRS/STIC 01N80/0502 and 01N80/0607 grants. }
\and
Francis Sourd%
\thanks{CNRS LIP6,
Universit\'e Pierre et Marie Curie,
Place de Jussieu, F-75005 Paris.
Francis.Sourd@lip6.fr.}
}

\maketitle

\begin{abstract}
  We study the problem of preemptive scheduling of $n$ equal-length
  jobs with given release times on $m$ identical parallel machines.
  The objective is to minimize the average flow time. Recently,
  Brucker and Kravchenko \cite{BRUCKKRAV} proved that the optimal
  schedule can be computed in polynomial time by solving a linear
  program with $O(n^3)$ variables and constraints, followed by some
  substantial post-processing (where $n$ is the number of jobs.) In
  this note we describe a simple linear program with only $O(mn)$
  variables and constraints. Our linear program produces directly the
  optimal schedule and does not require any post-processing.
\end{abstract}

\section{Introduction}

In the scheduling problem we study the input instance consists of $n$
jobs with given release times, where all jobs have the same processing
time $p$.  The objective is to compute a preemptive schedule of those
jobs on $m$ machines that minimizes the average flow time or,
equivalently, the sum of completion times, $\sum C_j$. In the standard
scheduling notation, the problem can be described as
$P|r_j,\pmtn,p_j=p|\sum C_j$.  Herrbach and Leung \cite{HL90} showed
that, for $m=2$, the optimal schedule can be computed in time $O(n
\log n)$. Du, Leung and Young \cite{DLY90} proved that the
generalization of this problem where processing times are arbitrary is
binary NP-hard.  We summarize these results in
Table~\ref{tab: complexity}.

Very recently, Brucker and Kravchenko \cite{BRUCKKRAV} gave a
polynomial-time algorithm for any number $m$ of machines.  Their
algorithm consists of two stages: first, they solve a complex linear
program with $O(n^3)$ variables and constraints, which is followed by
a post-processing stage where they construct an optimal schedule from
the optimal solution of this linear program.

\begin{table}[htb]
\begin{center}
\begin{tabular}{l|l}
        Problem & Complexity \\ \hline
  $P2 | r_j; \pmtn; p_j=p | \sum C_j$       
  & solvable in time $O(n\log n)$ \cite{HL90}    \\
  $P | r_j; \pmtn; p_j=p | \sum C_j$       
  & solvable in polynomial time \cite{BRUCKKRAV}, improved in this paper  \\
  $P | r_j; \pmtn\,\phantom{;p_j=p} | \sum C_j$       
  & binary NP-complete \cite{DLY90}            \\
  $P | \phantom{r_j;}\,\pmtn ; p_j=p | \sum C_j$       
  & solvable by the greedy algorithm (trivial)  \\
  $P | r_j; \phantom{\pmtn;}\,p_j=p | \sum C_j$       
  & solvable by the greedy algorithm (trivial)  \\
  $P2 | r_j; \pmtn;  p_j=p | \sum w_j C_j$       
  & open  \\
  $P | r_j; \pmtn; p_j=p | \sum w_j C_j$       
  & unary NP-complete \cite{LY90}  \\
\end{tabular}
\end{center}
\caption{Complexity of related scheduling problems.
$P2$ stands for the $2$-machine problem. In problems with the
objective function $\sum w_j C_j$, each job $j$ comes with a weight
$w_j$, and the goal is to minimize the weighted sum of all completion
times.}
\label{tab: complexity}
\end{table}
  
In this note, we show that there is always an optimal schedule in a
particular form, which we call \emph{normal}.  We then give a
simple linear program of size $O(mn)$, which directly
defines an optimal normal schedule.  As a side-product, we show that
there are optimal schedules with only $O(mn)$ preemptions, improving
the $O(n^3)$ bound on the number of preemptions from \cite{BRUCKKRAV}.

\section{Structural Properties}
\label{sec: structural properties}

\paragraph{Basic definitions.}
Throughout the paper, $n$ and $m$ denote, respectively, the number of
jobs and the number of machines. The jobs are numbered $1,2,\dots,n$
and the machines are numbered $1,2,\dots,m$.  All jobs have the same
length $p$.  For each job $j$, $r_j$ is the release time of $j$,
where, without loss of generality, we assume that $0 = r_1 \le \ldots
\le r_n$.

We define a \emph{schedule} $\calX$ to be a function which, for any
time $t$, determines the set $\calX(t)$ of jobs that are running at
time $t$. This set $\calX(t)$ is called the \emph{profile} at time
$t$. Let $\calX^{-1}(j)$ denote the set of times when $j$ is executed,
that is $\calX^{-1}(j) = \braced{t: j\in \calX(t)}$. In addition we
require that $\calX$ satisfies the following conditions:
\begin{description}
\item{(s1)} At most $m$ jobs are executed at any time, that is
  $|\calX(t)|\le m$ for all times $t$.
\item{(s2)} No job is executed before its release time, that is, for
  each job $j$, if $t < r_j$ then $j\notin \calX(t)$.
\item{(s3)} Each job runs in a finite number of time intervals.  More
  specifically, for each job $j$, $\calX^{-1}(j)$ is a finite union of
  intervals of type $[s,t)$.
\item{(s4)} Each job is executed for time $p$, that is
  $|\calX^{-1}(j)| = p$.
\end{description}
It is not difficult to see that condition (s3) can be relaxed to allow
jobs to be executed in infinitely (but countably) many intervals,
without changing the value of the objective function.

By $C_j = \sup\calX^{-1}(j)$ we denote the completion time of a job
$j$.  In this paper, we are interested in computing a schedule that
minimizes the objective function $\sum_{j=1}^n C_j$.

\smallskip

Note that, since we are dealing with preemptive schedules, it does not
matter to which specific machines the jobs in $\calX(t)$ are assigned
to.  When such an assignment is needed, we will use the convention
that the jobs are assigned to machines in the increasing order of
indices (or, equivalently, release times): the job with minimum index
is assigned to machine $1$, the second smallest job to machine $2$,
etc.


\paragraph{Some observations.}
We now show that, for the purpose of minimizing our objective
function, we can restrict our attention to schedules with some
additional properties.

Call a schedule $\calX$ \emph{left-adjusted} if it satisfies the
following condition: for any times $s < t$, where $|\calX(s)| < m$, if
$j\in \calX(t)$ and $r_j\le s$ then $j\in\calX(s)$ as well.  Any
optimal schedule is left-adjusted, for otherwise, if the above
condition is not satisfied, we can move a sufficiently small portion
$\epsilon > 0$ of $j$ from the last block where it is executed to the
interval $[s,s+\epsilon)$, obtaining a feasible schedule in which the
completion time of $j$ is reduced by $\epsilon$ and other completion
times do not change. Thus we only need to be concerned with
left-adjusted schedules.

We say that the \emph{completion times are ordered} in a schedule
$\calX$, if $C_1\le C_2\le \ldots \le C_n$.  Brucker and Kravchenko
\cite{BRUCKKRAV} showed that any schedule can be converted into one
with ordered completion times, without increasing the objective value.


\paragraph{Irreducible schedules.} 
We say that a schedule $\calX$ is \emph{irreducible} if it is
left-adjusted and satisfies
the following condition for any times $s < t$:
\begin{eqnarray}
        \max(\,\calX(s) - \calX(t)\,) &<& \min(\, \calX(t) - \calX(s)\,).
        \label{eqn: irreducible}
\end{eqnarray}
In the formula above we use the convention
that $\max(\emptyset) = -\infty$ and $\min(\emptyset) = +\infty$,
so (\ref{eqn: irreducible}) holds whenever
$\calX(s)\subseteq \calX(t)$ or $\calX(t)\subseteq \calX(s)$.

For a schedule $\calX$ and a job $j$, define the \emph{halfway point}
of $j$ as\footnote{This is a standard value in scheduling, even though
  the factor $\onehalf$ is irrelevant for this paper.}  $H_j(\calX) =
\onehalf \int_{\calX^{-1}(j)} t \, \mathrm{d}t$.  Then let $H(\calX) =
(H_1(\calX), H_2(\calX),\ldots,H_n(\calX))$.  For two different
schedules $\calX$, $\calY$, we say that $H(\calX)$ is
\emph{lexicographically smaller} than $H(\calY)$, if $H_i(\calX) <
H_i(\calY)$ for the smallest $i$ for which $H_i(\calX) \neq
H_i(\calY)$.

\begin{lemma}\label{lem: exchange}
Let $\calX$ be a schedule and $[s,s+\epsilon)$, $[t,t+\epsilon)$
two time intervals such that $s+\epsilon\le t$, for some $\epsilon > 0$.
Suppose that there are two jobs $i<j$ with $r_j\le s$, such that
$[s,s+\epsilon)\subseteq \calX^{-1}(j)-\calX^{-1}(i)$
and $[t,t+\epsilon) \subseteq \calX^{-1}(i) - \calX^{-1}(j)$.
Let $\calY$ denote the schedule obtained from $\calX$ by exchanging
jobs $i,j$ in intervals $[s,s+\epsilon)$, $[t,t+\epsilon)$.
Then $H(\calY)$ is lexicographically strictly smaller than $H(\calX)$.
\end{lemma}

\begin{proof}
Clearly, $H_i(\calY) = H_i(\calX) - (t-s)\epsilon/2$, and
$H_k(\calY)=H_k(\calX)$ for all $k<i$. This directly implies the lemma.
\end{proof}

We need to show that there exists an optimal irreducible schedule.
This is quite easy to show if we put some restrictions on the granularity of the
schedules, for example if we assume that schedules are constant in
each unit interval $[t,t+1)$ for $t\in\mathbb N$.  In that case one can
show that after finite number of exchanges (as defined in the previous
lemma) any optimal schedule can be transformed into an irreducible
optimal schedule. This applies, in particular, to the case when
the processing time and all release times are integer \cite[Theorem 6]{BRUCKKRAV}. 
The proof for arbitrary real numbers is more difficult, and is based
on the following lemma, whose proof appears in the appendix.


\begin{lemma}\label{lem: lexicographically}
There exists an optimal schedule $\calX$ for which the
vector $H(\calX)$ is lexicographically minimum over all optimal schedules.
\end{lemma}

\begin{proof}
See Appendix~\ref{sec: irreducible general case}.
\end{proof}


\begin{lemma}\label{lem: normal}
There exists an optimal schedule that is irreducible.
\end{lemma}

\begin{proof} 
  Let $\calX$ be an optimal schedule which minimizes $H(\calX)$ among
  all optimal schedules. According to Lemma~\ref{lem:
  lexicographically}, $\calX$ is well defined.  By optimality, $\calX$
  is left-adjusted.

  We claim that the completion times in $\calX$ are ordered.  Towards
  contradiction, suppose there are jobs $i<j$ with $C_i>C_j$.  Let
  $[t,t+\epsilon) \subseteq [C_j,C_i)$ be an interval where $i$ is
  scheduled and $[s,s+\epsilon) \subseteq [0,C_j)$ a arbitrary
  interval, where $j$ but not $i$ is scheduled.  This contradicts the
  minimality of $H(\calX)$ by Lemma~\ref{lem: exchange}.
  
  We claim that $\calX$ also satisfies (\ref{eqn: irreducible}).
  Towards contradiction, suppose it does not.  Then there are time
  intervals $[s,s+\epsilon)$ and $[t,t+\epsilon)$ for $s+\epsilon\le
  t$ and jobs $i < j$ such that $\calX$ schedules $j$ but not $i$ in
  $[s,s+\epsilon)$ and schedules $i$ but not $j$ in $[t,t+\epsilon)$.
  Then Lemma~\ref{lem: exchange} applies as before and the proof is
  now complete.
\end{proof}

\medskip

We now give a characterization of irreducible schedules that will play
a major role in the construction of our linear program.

For a given job $j$ and a time $t$ we partition $\calX(t) -
\braced{j}$ into jobs released earlier and jobs released later than
$j$.  Formally, $\calXbef{j}(t) = \braced{i\in \calX(t): i<j}$ and
$\calXaft{j}(t) = \braced{i\in \calX(t): i>j}$, see figure~\ref{fig:
  normal structure}.  The lemma below provides a characterization of
irreducible schedules.

\begin{figure}[htb]
\centerline{\input{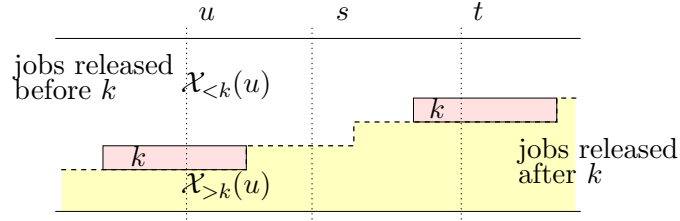}}
\caption{Structure of any irreducible schedule.}
\label{fig: normal structure}
\end{figure}


\begin{lemma}\label{lem: order}
Let $\calX$ be an irreducible schedule.
Suppose that we have two times $u < t$ and a job $k$ such
that $r_k\le u$. Then:

\noindent{\rm (a)}
If $k\in\calX(u) - \calX(t)$ then
$|\calXbef{k}(t)| \le |\calXbef{k}(u)|$.

\noindent{\rm (b)}
If $k\in\calX(t) - \calX(u)$ then
$|\calX(u)| = m$,  
$|\calXaft{k}(t)| \ge |\calXaft{k}(u)|$, and
$|\calXbef{k}(t)| < |\calXbef{k}(u)|$.

\noindent{\rm (c)}
If $k\in\calX(u) \cap \calX(t)$ then
$|\calXbef{k}(t)| \le |\calXbef{k}(u)|$ and
$|\calXaft{k}(t)| \ge |\calXaft{k}(u)|$.
\end{lemma}

\begin{proof}
(a)
If there was a $j\in \calXbef{k}(t) - \calXbef{k}(u)$,
this would imply that
$\max(\calX(u) - \calX(t)) \ge k
         > j \ge \min(\calX(t) - \calX(u))$,
contradicting irreducibility. Thus (a) follows.

(b)
Since $k\in\calX(t) - \calX(u)$ and $r_k \le u$, the assumption that
$\calX$ is left-adjusted implies that $|\calX(u)| = m$.

We must have $\calXaft{k}(u) \subseteq \calXaft{k}(t)$, for otherwise,
the existence of $k\in \calX(t) - \calX(u)$ and an $l \in
\calXaft{k}(u) - \calXaft{k}(t)$ would contradict irreducibility.  The
inequality $|\calXaft{k}(u)| \le |\calXaft{k}(t)|$ follows.  This, the
assumption of the case, and $|\calX(u)| = m$ imply $|\calXbef{k}(u)| >
|\calXbef{k}(t)|$.

(c) We only prove the first inequality, as the proof for the second
  one is very similar.  Towards contradiction, suppose
  $|\calXbef{k}(u)| < |\calXbef{k}(t)|$, and pick any
  $i\in\calXbef{k}(t) - \calXbef{k}(u)$.  Then $r_i\le r_k\le u$ and
  $i\in \calX(t)-\calX(u)$, and so the assumption that $\calX$ is
  left-adjusted implies $\calX(u) = m$.  This, in turn, implies that
  $|\calXaft{k}(u)| > |\calXaft{k}(t)|$, so we can choose $j\in
  \calXaft{k}(u) - \calXaft{k}(t)$.  But this means that $j > k > i$
  and $j\in \calX(u) - \calX(t)$, and the existence of such $i$ and
  $j$ contradicts irreducibility.
\end{proof}

\section{A Simple Linear Program}

\paragraph{Machine assignment.}
We now consider the actual job-machine assignment in an irreducible
schedule $\calX$.  As explained earlier, at every time $t$ we assign
the jobs in $\calX(t)$ to machines in order, that is job $j\in
\calX(t)$ is assigned to machine $1+|\calXbef{j}(t)|$.
Lemma~\ref{lem: order} implies that, for any fixed $j$, starting
at $t=r_j$ the value of
$|\calXbef{j}(t)|$ decreases monotonically with $t$. Therefore, with
machine assignments taken into account, $\calX$ will have the
structure illustrated in Figure~\ref{fig: normal structure}.

Call a schedule $\calX$ \emph{normal} if for each job $j$ and each
machine $q$, job $j$ is executed on $q$ in a single (possibly empty)
interval $[C_{j,q},S_{j,q})$, and
\begin{description}
\item{(1)} 
$C_{j,q}\le S_{j+1,q}$ for each machine $q$ and job $j<n$, and
\item{(2)}
$C_{j,q}\le S_{j,q-1}$ for each machine $q>1$ and job $j$.
\end{description}

\begin{figure}[htb]
\centerline{\epsfig{file=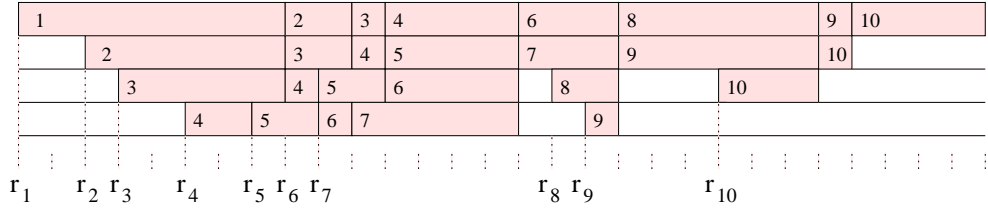,width=13cm}}
\caption{Example of a normal schedule. The processing
        time is $p=8$.}
\label{fig: normal example}
\end{figure}

By the earlier discussion, each irreducible schedule is normal
(although the reverse does not hold.)  An example of a normal (and
irreducible) schedule is shown in Figure~\ref{fig: normal example}.

\paragraph{Linear program.}
We are now ready to construct our linear program:
\begin{align}
{\mbox{\rm minimize }}   && \textstyle{\sum_{j=1}^n C_{j,1}}& 
                \label{eqn: linear program} \\
{\mbox{\rm subject to}} &&
           - S_{j,m}              \; &\le - r_j      && j = 1,\dots,n
                        \nonumber \\
        && \sum_q (C_{j,q} - S_{j,q})\; &= p         && j = 1,\dots,n
                          \nonumber \\
        && S_{j,q}   - C_{j,q}     \; &\le 0          && j = 1,\dots,n,\;  q = 1,\dots, m
                        \nonumber \\
        && C_{j,q} - S_{j,q-1}   \; &\le 0          && j = 1,\dots, n,\; q = 2,\dots,m
                        \nonumber \\
        && C_{j,q} -S_{j+1,q}  \; &\le 0          && j = 1,\dots,n-1,\;  q = 1,\dots,m
                        \nonumber
\end{align}


The correspondence between normal schedules and feasible solutions to
this linear program should be obvious. For any normal schedule, the
start times $S_{j,q}$ and completion times $C_{j,q}$ satisfy the
constraints of (\ref{eqn: linear program}).  And vice versa, for any
set of the numbers $S_{j,q}$, $C_{j,q}$ that satisfy the constraints
of (\ref{eqn: linear program}), we get a normal schedule by scheduling
any job $j$ in interval $[S_{j,q},C_{j,q})$ on each machine $q$.  Thus
we can identify normal schedules $\calX$ with feasible solutions of
(\ref{eqn: linear program}).  Note, however, that in $\calX$ a job $j$
could complete earlier than $C_{j,1}$ (this can happen when $C_{j,1} =
S_{j,1}$.) Thus the only remaining issue is whether the optimal normal
schedules correspond to optimal solutions of (\ref{eqn: linear
  program}).


\begin{theorem} 
  The linear program above correctly computes an optimal schedule.
  More specifically, $\min_\calX\sum_jC_j = \min_\calX\sum_jC_{j,1}$,
  where the minima are over normal schedules $\calX$, and $C_j$
  represents the completion time of job $j$ in $\calX$.
\end{theorem}

\begin{proof}
  By the correspondence between normal schedules and feasible
  solutions of (\ref{eqn: linear program}), discussed before the
  theorem, we have $C_j\le C_{j,1}$ for all $j$, and thus the $\le$
  inequality is trivial.
  
  To justify the other inequality, fix an optimal irreducible (and
  thus also normal) schedule $\calX$ and a job $j$, and let
  $[S_{j,q},C_{j,q})$ be the last (that is, the one with minimum $q$)
  non-empty execution interval of $j$.  Consider a block $[s,t)$ where
  $t=C_{j,q}$.  By Lemma~\ref{lem: order} all jobs executed on
  machines $1,2,\dots,q-1$ in $[s,t)$ are numbered lower than $j$.
  Further, by the ordering of completion times, they are not executed
  after $t$. Thus they must be completed at $t$ as well. Therefore we
  can set $[S_{j,h},C_{j,h})=[t,t)$, for all machines $1\le h<q$,
  without violating any inequality. This gives a normal schedule in
  which $C_j = C_{j,1}$.
\end{proof}

\section{Final Remarks} \label{sec:more}

We proved that the scheduling problem $P|r_j,\pmtn,p_j=p|\sum C_j$
can be reduced to solving a linear program with
$O(mn)$ variables and constraints. This leads to a polynomial
time algorithm more efficient than the one resulting from \cite{BRUCKKRAV}.
The question whether linear programming can be avoided, and whether
this problem can be solved with a combinatorial,
strongly polynomial time algorithm (whose number of steps is
a polynomial function of only $m$ and $n$) remains open. 
Our characterizations of optimal schedules could be helpful
in designing such an algorithm.

Brucker and Kravchenko \cite{BRUCKKRAV} showed that there is an
optimal schedule with $O(n^3)$ preemptions. Our proof provides a
better, $O(mn)$ bound on the number of preemptions, since in an
irreducible schedule each job is preempted at most $m-1$ times.  (Of
course, we can always assume that $m\le n$.) We do not know whether
this bound is asymptotically tight.  It is thus quite possible that
there exist optimal schedules in which the number of preemptions is
$O(n)$, independent of $m$. If this is true, this could lead to
efficient combinatorial algorithms for this problem whose running time
is even independent of $m$, perhaps even as fast as $O(n\log n)$.  Such
improvement would require a much deeper study of the structural
properties of optimal schedules.  Since we use only the existence of
normal optimal schedules, rather than irreducible schedules, we feel
that the problem has more structure to be exploited.

We implemented the complete algorithm (converting the instance to a
linear program and solving this linear program). It is accessible at
{\verb#http://www.lri.fr/~durr/P_rj_pmtn_pjp_sumCj#}.

\bibliographystyle{plain}
\bibliography{sched}

\appendix

\section{Proof of Lemma~\ref{lem: lexicographically}}
\label{sec: irreducible general case}

\newtheorem{lemma_new}{Lemma}
\begin{lemma_new}
There exists an optimal schedule $\calY^\ast$ that minimizes
lexicographically $H(\calY^\ast)$ among all optimal schedules.
\end{lemma_new}

\begin{proof}
  The major difficulty that we need to overcome is that the set of
  schedules is not closed as a topological space, so there could be a
  sequence of schedules with decreasing values of $H(\calX)$ whose
  limit is not a legal schedule. The idea of the proof is to reduce
  the problem to minimizing $H(\calX)$ over a compact subset of
  schedules.

  Define a \emph{block} of a schedule $\calX$ to be a
  maximal time interval $[u,t)$ such that $(u,t)$ does not contain any
  release times and $\calX(s)$ is constant for $s\in[u,t)$.
  
  For convenience, let $r_{n+1}$ to be any upper bound on the last
  completion time of any optimal schedule, say $r_{n+1} = r_n+ np$.
  Thus all jobs are executed between $r_1$ and $r_{n+1}$.  Each
  interval $[r_i,r_{i+1})$, for $i=1,\dots,n$ is called a
  \emph{segment}.  By condition (s3), each segment is a disjoint union
  of a finite number of blocks of $\calX$.  Also, for each job $j$, we
  have $C_j = t$ for the last non-empty block $[s,t)$ whose profile
  contains $j$.
  
  A schedule $\calX$ is called \emph{tidy} if all jobs are completed
  no later than at $r_{n+1}$ and, for any segment $[r_i,r_{i+1})$, the
  profiles $\calX(t)$, for $t\in [r_i,r_{i+1})$, are lexicographically
  ordered from left to right. More precisely, this means that, for any
  $r_i \le s < t < r_{i+1}$, we have
\begin{eqnarray*}
        \min(\,\calX(s) - \calX(t)\,) &\le& \min(\, \calX(t) - \calX(s)\,).
\end{eqnarray*}
One useful property of tidy schedules is that its total number of
blocks (including the empty ones) is $N = nM$, where $M = \sum_{l =
  0}^m\binom{n}{l}$.  From now on we identify any tidy schedule
$\calX$ with the vector $\calX\in \reals^N$ whose $b$-th coordinate
$x_b$ represents the length of the $b$-th block in $\calX$.

In fact, the set $\bfT$ of tidy schedules is a (compact) convex
polyhedron in $\reals^N$, for we can describe $\bfT$ with a set of
linear inequalities that express the following constraints:
\begin{itemize}
\item Each job $j$ is not executed before $r_j$,
\item Each job $j$ is executed for time $p$.
\end{itemize}
For example, the second constraint can be written as $\sum_b x_b = p$,
where the sum is over all blocks $b$ whose profile contains $j$.

\begin{claim} \label{cla: completion ordering}
{\rm \cite{BRUCKKRAV}}\,
Any schedule $\calX$ can be transformed into a schedule
with ordered completion times, without increasing the
objective function value.
\end{claim}

  Suppose for jobs $i<j$ the completion times in $\calX$ satisfy
  $C_i>C_j$.  Then there must be a maximal time $t$, such that in
  $[t,C_i)$ both jobs are scheduled for an equal amount of time.
  Exchanging both jobs in this interval will reorder their completion
  times. After repeating this process sufficiently many times, eventually
  all completion times will be ordered. See~\cite{BRUCKKRAV} for
  details.

\begin{claim}\label{cla: tidy schedules}
  Let $\calX$ be a schedule in which completion times are ordered and
  upper bounded by $r_{n+1}$.  Then $\calX$ can be converted into a
  tidy schedule $\calX'$ such that

\noindent {\rm (a)} $C'_j \le C_j$ for all $j$ (where $C_j$ and $C'_j$
        are the completion times of $j$ in $\calX$ and $\calX'$,
        respectively.) 

\noindent {\rm (b)} $H(\calX')$ is equal to or lexicographically
         smaller than $H(\calX)$.
\end{claim}

Indeed, suppose that $\calX$ has two consecutive blocks $A = [u,s)$,
$B = [s,t)$ where $r_i\le u < s < t \le r_{i+1}$, and the profile
$\calX(u)$ of $A$ is larger (lexicographically) than the profile
$\calX(s)$ of $B$.  Exchange $A$ and $B$, and denote by $\calX'$ the
resulting schedule.  Let $j = \min(\calX(s)-\calX(u)) <
\min(\calX(u)-\calX(s))$.  Since $C_j \ge t$, all jobs in
$\calX(u)-\calX(s)$ are also completed not earlier than at $t$. So
this exchange does not increase any completion times.  We have
$H_j(\calX') < H_j(\calX)$ and $H_i(\calX') = H_i(\calX)$ for $i<j$.
Thus $H(\calX')$ is lexicographically smaller than $H(\calX)$.  By
repeating this process, we eventually convert $\calX$ into a tidy
schedule that satisfies the claim.

\smallskip

We now continue the proof of the lemma.  Fix some optimal schedule
$\calX^\ast$.  Let $C^\ast_j$ denote the completion time of a job $j$
in $\calX^\ast$.  From Claims~\ref{cla: completion ordering} and
\ref{cla: tidy schedules}, we can assume that $\calX^\ast$ is tidy and
$C^\ast_1 \le C^\ast_2 \le \ldots \le C^\ast_n$.  (For the peace of
mind, it is worth noting that Claim~\ref{cla: tidy schedules} implies
that $\calX^\ast$ is well defined, for it reduces the problem to
minimizing $\sum_j C_j$ over a compact subset $\bfT$ of $\reals^N$.)
 
Consider a class $\bfT_0 \subseteq \bfT$ of tidy schedules $\calX$
such that each job $j$ in $\calX$ is completed not later than
$C^\ast_j$. Since $\calX^\ast\in\bfT_0$, the set $\bfT_0$ is not
empty.  Similarly as $\bfT$, $\bfT_0$ is a (compact) convex
polyhedron.  Indeed, we obtain $\bfT_0$ by using the same constraints
as for $\bfT$ and adding the constraints that each job $j$ is
completed not later than at $C^\ast_j$.  To express this constraint,
if in $\calX^\ast$ the completion time $C^\ast_j$ of $j$ is at the end
of the $a$-th block in the segment $[r_i,r_{i+1})$, then for each $b >
(i-1)M+a$ such that $j$ is in the profile of the $b$-th block, we
would have a constraint $x_b = 0$.  Note that these constraints do not
explicitly force $j$ to end \emph{exactly} at $C^\ast_j$, but the
optimality of $\calX^\ast$ guarantees that it will have to.

Now we show that there exists a schedule $\calY^\ast \in\bfT_0$ for
which $H(\calY^\ast)$ is lexicographically minimum.  First, as we
explained earlier, $\bfT_0$ is a compact convex polyhedron.  Let
$\bfT_1\subseteq \bfT_0$ be the set of $\calX$ for which $H_1(\calX)$
is minimized.  $H_1$ is a continuous quadratic function over $\bfT_0$,
and thus $\bfT_1$ is also a non-empty compact set.  Continuing this
process, we construct sets $\bfT_2, \ldots, \bfT_n$, and we choose
$\calY^\ast$ arbitrarily from $\bfT_n$.
\end{proof}

\end{document}